\documentclass[journal=ancac3,manuscript=article,layout=twocolumn]{achemso}

\usepackage[version=3]{mhchem} 
\usepackage{bm}
\usepackage[]{graphicx}
\usepackage{color}

\usepackage{amsmath,amssymb}
\usepackage[tight]{units}
\usepackage[colorlinks=true,citecolor=blue,linkcolor= blue,linkcolor=blue,urlcolor=blue]{hyperref}

\newcommand{\SbTe}{$\text{Sb}_2\text{Te}_3$~}
\newcommand{\f}[1]{Fig.~\ref{fig:#1}}
\newcommand{\fs}[1]{Figs.~\ref{fig:#1}}

\author{Nicki F. Hinsche}
\email{nicki.hinsche@physik.uni-halle.de}
\affiliation{Institute of Physics, Martin Luther University Halle-Wittenberg, D-06099 Halle, Germany}
\author{Sebastian Zastrow}
\affiliation{Institute of Nanostructure and Solid State Physics, Universit{\"a}t Hamburg, Jungiusstrasse 11, D-20355 Hamburg, Germany}
\author{Johannes Gooth}
\affiliation{Institute of Nanostructure and Solid State Physics, Universit{\"a}t Hamburg, Jungiusstrasse 11, D-20355 Hamburg, Germany}
\author{Laurens Pudewill}
\affiliation{Institute of Nanostructure and Solid State Physics, Universit{\"a}t Hamburg, Jungiusstrasse 11, D-20355 Hamburg, Germany}
\author{Robert Zierold}
\affiliation{Institute of Nanostructure and Solid State Physics, Universit{\"a}t Hamburg, Jungiusstrasse 11, D-20355 Hamburg, Germany}
\author{Florian Rittweger}
\affiliation{Max Planck Institute of Microstructure Physics, Weinberg 2, D-06120 Halle, Germany}
\author{Tom\'{a}\v{s} Rauch}
\affiliation{Institute of Physics, Martin Luther University Halle-Wittenberg, D-06099 Halle, Germany}
\author{J\"{u}rgen Henk}
\affiliation{Institute of Physics, Martin Luther University Halle-Wittenberg, D-06099 Halle, Germany}
\author{Kornelius Nielsch}
\affiliation{Institute of Nanostructure and Solid State Physics, Universit{\"a}t Hamburg, Jungiusstrasse 11, D-20355 Hamburg, Germany}
\author{Ingrid Mertig}
\affiliation{Institute of Physics, Martin Luther University Halle-Wittenberg, D-06099 Halle, Germany}
\alsoaffiliation{Max Planck Institute of Microstructure Physics, Weinberg 2, D-06120 Halle, Germany}
\title[]
  {Impact of the Topological Surface State on the Thermoelectric Transport in Sb$_2$Te$_3$ Thin Films}

\abbreviations{}

\begin{document}


\begin{abstract}
\textit{Ab initio} electronic structure calculations based on density functional theory and tight-binding methods 
for the thermoelectric properties of $p$-type \SbTe films are presented. The thickness-dependent 
electrical conductivity and the thermopower are computed in the diffusive limit of transport based on the Boltzmann equation. 
Contributions of the bulk and the surface to the transport coefficients are separated which enables to identify a clear impact 
of the topological surface state on the thermoelectric properties. By tuning the charge carrier concentration, 
a crossover between a surface-state-dominant and a Fuchs-Sondheimer transport regime is achieved. 
The calculations are corroborated by thermoelectric transport measurements on \SbTe films grown by atomic layer deposition. 
\end{abstract}

%
%
Almost all proposed three-dimensional (3D) ${Z}_2$ topological insulators (TIs) \cite{Zhang:2009p14347,Ando:2013p16362} are efficient thermoelectric materials. That is not by coincidence, since the link between an efficient thermoelectric material and the topological character is the inverted band gap \cite{Kong:2011p16550,Rittweger:2014p16422}. 
The last is due to spin-orbit coupling which switches parity of the bands and leads, if strong enough, to narrow band gaps which are favourable for efficient room-temperature thermoelectrics. Usually strong spin-orbit coupling is 
mediated by heavy elements which in turn also tend to reduce the material's lattice thermal conductivity, another requirement for desirable thermoelectrics. 

In the early 90's, Hicks and Dresselhaus \cite{Hicks:1993p14911,Hicks:1993p15780} proposed the concept of low-dimensionality to increase further the thermoelectric efficiency; primarily in thin films, the thermopower $S$ should be enlarged. 
However, in contrast to previous theoretical model calculations \cite{Ghaemi:2010,Takahashi:2012}, decreased values of $S$ were found experimentally  \cite{Boulouz:2001p16385,Peranio:2006p15247,Zastrow:2014p16429} for \ce{Bi2Te3} and \ce{Sb2Te3} thin films and were recently corroborated by both model \cite{Osterhage:2014p16634,Gooth:2014p16760} and \textit{ab initio} calculations \cite{Rittweger:2014p16422} of our groups. 

Thus, to resolve this discrepancy, the potential impact of the surface state (SS) of TIs on thermoelectricity needs to be investigated in more detail. In this Letter, we present \textit{ab initio} calculations and transport measurements of the thermoelectric properties of \SbTe films at varying thickness, temperature and charge carrier concentration. 

\subsection{Theoretical results}
\begin{figure*}[t]
\centering
\includegraphics[width=0.7\textwidth]{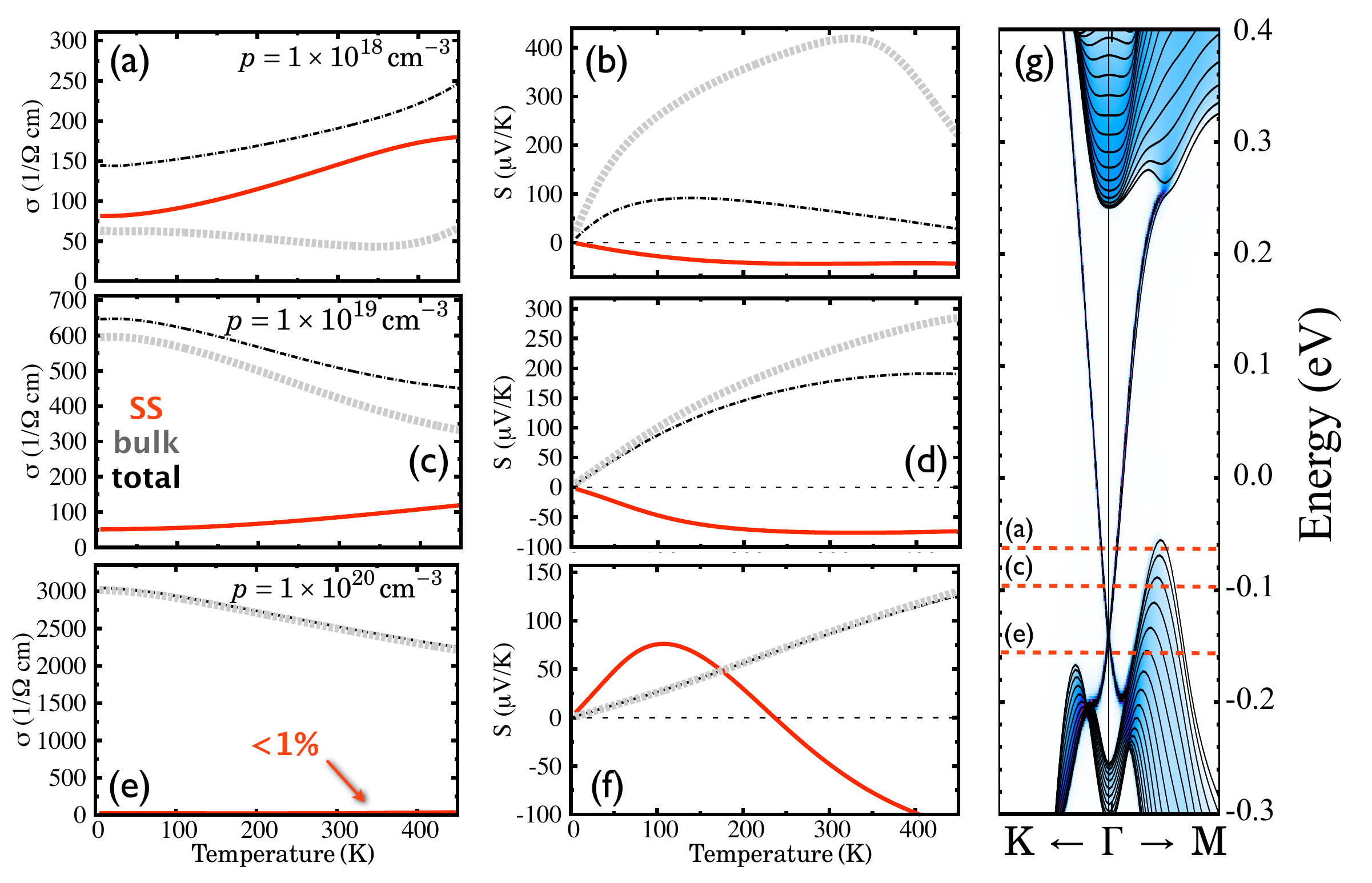}
  \caption{(color online) Electrical conductivity (a),(c),(e) and thermopower (b),(d),(f) in dependence on temperature for three distinct 
hole concentrations at a film thickness of 18 QL. Hole concentration $n = \unit[1 \times 10^{18}]{cm^{-3}}$ in (a) and (b), $p = \unit[1 \times 10^{19}]{cm^{-3}}$ in (c) and (d), $p = \unit[1 \times 10^{20}]{cm^{-3}}$ in (e) and (f). Pure bulk contributions are represented by gray dashed lines, the contribution of the surface 
states is given by red solid lines, while black dash-dotted lines show the total contribution of the sample. In (g) the band structure of the 18 QL 
\SbTe film is shown around the fundamental band gap. Thin red dashed lines indicate the position of the chemical potential $\mu$ at zero temperature.}
\label{fig:temp}
\end{figure*}
In the following, we discuss the doping- and temperature-dependent electrical conductivity and 
thermopower, as shown in \f{temp}, exemplary for a \SbTe film thickness of 18 quintuple-layer (QL), \textit{i.e.} about $\unit[18]{nm}$. 
A discussion of films with other thicknesses is given in the supplemental material \cite{supp}.

The converged electronic structure results serve as input to obtain the thermoelectric transport properties 
at temperature $T$ and fixed extrinsic charge carrier concentration $p$ 
by solving the linearized Boltzmann equation in relaxation time approximation (RTA) \cite{Mertig:1999p12776}. 
By using special projection techniques \cite{Rittweger:2014p16422} we distinguish between contributions 
from bulk states (gray dashed lines), SSs (red solid lines) and the total contribution (black dash-dotted line), 
defined as $\sigma_{\text{tot}}=\sigma_{\text{bulk}}+\sigma_{\text{SS}}$ and $S_{\text{tot}}=\nicefrac{(\sigma_{\text{bulk}}S_{\text{bulk}}+\sigma_{\text{SS}}S_{\text{SS}})}{\sigma_{\text{tot}}}$. 
Three typical charge carrier concentrations are chosen to reflect the overall behavior of the transport properties. 
Only $p$-type thin films will be discussed because \SbTe is inherently $p$-type conductive \cite{Thonhauser:2003p15461,Jiang:2012p16559} due to intrinsic defects, \textit{i.e.} Sb vacancies and Sb$_{\text{Te}}$ antisites.
 
Due to the fact that a 3D TI offers robust metallic\footnote{We denote the SS metallic, as it shows clear transport signatures of a conventional metal.}
SSs within the insulating bulk band
gap, an enhanced electrical conductivity of the entire system compared to a conventional insulator 
is expected for very small charge carrier concentrations, \textit{i.e.},
if the chemical potential is situated in or nearby the bulk band gap. The latter scenario is shown in \f{temp}(a) 
for a $p$-type doping of $p = \unit[1 \times 10^{18}]{cm^{-3}}$. 
At low temperatures the contribution $\sigma_{\text{SS}}$ of the SS (red solid line) is already larger 
than the bulk contribution $\sigma_{\text{bulk}}$. 
With increasing temperature the chemical potential shifts into the band gap, thereby decreasing $\sigma_{\text{bulk}}$ until bipolar conduction contributes at 
about $T=\unit[300]{K}$. In contrast $\sigma_{\text{SS}}$ increases monotonically with temperature. 
The latter can easily be understood, having in mind that for a two-dimensional system, \textit{i.e.} the SS, 
the transport distribution function (TDF) scales as $\Sigma(\mu) \propto \text{d}l_{\text{F}}\times v$, 
where $\text{d}l_{\text{F}}$ is the circumference of the Fermi circle at chemical potential $\mu$. 
For energies close to the Dirac point $v_\text{SS}$ is constant in energy, while $\text{d}l_{\text{F}}\propto E$. 
Consequently, the TDF is linear in energy. 
Small deviations from the latter arise at about $E \approx \unit[0.1]{eV}$ and are 
attributed to the hexagonal warping of the Fermi surface. 

The impact of the SS on the total thermopower, as shown in \f{temp}(b), is even more remarkable. 
While the bulk contribution $S_{\text{bulk}}$ shows the typical behavior for a $p$-type narrow-band gap semiconductor 
with a maximum of $S_{\text{bulk}} \approx \unit[410]{\mu V/K}$ at $T=\unit[325]{K}$, the contribution of the SS shows 
the expected metallic behavior with small values of $S_{\text{SS}} \le -\unit[50]{\mu V/K}$. 
The negative sign stems from the mere fact that the SS above (below) the Dirac point mimics the slope of 
the band dispersion of conduction (valence) bands. 

Summarizing, the influence of the SS leads to a clearly diminished total thermopower (black dash-dotted line). 
As a rule of thumb, assuming $S_{\text{bulk}} \gg S_{\text{SS}}$ yields 
$S_{\text{tot}} \approx \nicefrac{S_{\text{bulk}}}{(\eta+1)}$ with $\eta = \nicefrac{\sigma_{\text{SS}}}{\sigma_{\text{bulk}}}$. 
The larger the contribution of the SS to the total electrical conductivity, the smaller the total thermopower $S_{\text{tot}}$. 

The value of the total thermopower ($\unit[65]{\mu V/K}$ at $\unit[300]{K}$; black dashed-dotted line in \f{temp}(b)) is 
reduced to less than one fifth of the bulk value, which corroborates 
the above-noted estimation. We note that, within the RTA, the signature of the SS on $S_{\text{tot}}$ relies 
to a certain extent on the ratio $\nicefrac{\tau_{\text{SS}}}{\tau_{\text{bulk}}}$ of the relaxation times. 
If $\tau_{\text{bulk}} \gg \tau_{\text{SS}}$ the reduction of the total thermopower due to the conducting 
SS would be much weaker than proposed. With regard to the backscattering protection of the SS, however, the opposite scenario $\tau_{\text{SS}} \gg \tau_{\text{bulk}}$ should be expected \cite{Roushan:2009p16199}.

For thermoelectric applications charge carrier concentration of $p = \unit[1 \times 10^{19}]{cm^{-3}}$ (cf. \fs{temp}(c) and (d)) 
are more applicable. Here the contribution of the SS to $\sigma_{\text{tot}}$ is smaller, compared to the low doped case, as 
the chemical potential is located closer to the Dirac point (cf. \f{temp}(c)). Still, a reduction of the film's total thermopower by 45\% to 
$S_{\text{tot}} \approx \unit[187]{\mu V/K}$ at $T=\unit[400]{K}$ compared to the bare bulk value 
is calculated due to the impact of the metallic SS (cf. \f{temp}(d)).

In heavy $p$-doped samples, at $p = \unit[1 \times 10^{20}]{cm^{-3}}$, as depicted in \fs{temp}(e) and (f), 
the absolute value of $S_{\text{SS}}$ still reaches about $\unit[25]{(\Omega cm)^{-1}}$ at room temperature but 
is negligibly small compared to the bulk contribution. This is reflected in the thermopower, 
for which $S_{\text{tot}} \approx S_{\text{bulk}}$. Focussing on the SS contribution the picture is more delicate. At low 
temperatures, $S_{\text{SS}}$ is positive but decreases at elevated temperatures and changes 
sign at about $T=\unit[250]{K}$. The latter is attributed to a shift of the chemical potential above the Dirac point, 
where the slope of both the TDF and $S_{\text{SS}}$ change sign. 



\begin{figure*}[t!]
\centering
\includegraphics[width=0.75\textwidth]{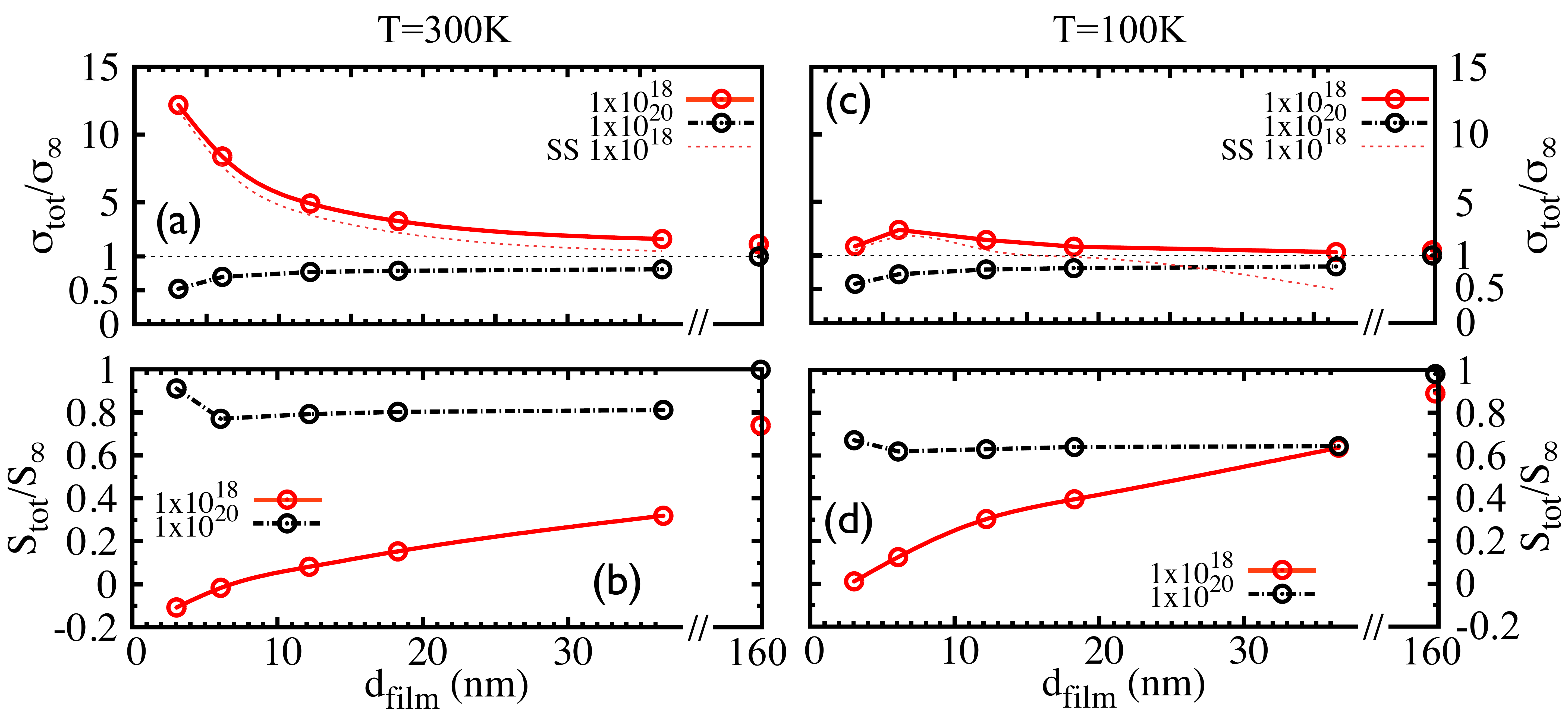}
  \caption{(color online) Thickness-dependent thermoelectric transport properties. Total electrical conductivity and thermopower at 
  $T=\unit[300]{K}$ (a and b) and $T=\unit[100]{K}$ (c and d), respectively. 
  The results are depicted for two distinct hole concentrations; $p = \unit[1 \times 10^{18}]{cm^{-3}}$ (red solid lines) and $p = \unit[1 \times 10^{20}]{cm^{-3}}$ (black dash-dotted lines). For the electrical conductivity in (a) and (c) the pure contribution of the surface state is given additionally as a thin dashed red line. All the transport properties are normalized to the bulk values $\sigma_{\infty}$ and $S_{\infty}$.}
  \label{fig:thick}
\end{figure*}
After discussing the temperature and doping-dependent thermoelectric transport properties for a film of 18 QL thickness, 
we now focus on the film thickness dependence of the transport properties at $T=\unit[300]{K}$ and $T=\unit[100]{K}$, as shown 
in \f{thick}. At room-temperature, two distinct transport regimes can be discussed with respect to the dependence of 
the charge carrier concentration of the films. At heavy $p$-doping ($p = \unit[1 \times 10^{20}]{cm^{-3}}$, black dash-dotted lines in \f{thick}), 
the film behaves almost metallic without significant influence of the topological SS on the transport properties. 
With increasing film thickness the normalized electrical conductivity at room temperature (cf. \f{thick}(a)) tends asymptotically to the bulk value as 
$\nicefrac{\sigma_{\text{tot}}}{\sigma_{\infty}} \sim (1+(\nicefrac{3v\tau}{8d}))^{-1}$ in accordance with the Fuchs-Sondheimer theory \cite{Fuchs:1938p16454,Sondheimer:1952p16552}. While derived to 
describe surfaces the Fuchs-Sondheimer theory does not account for the influence of SS's and quantum size effects. 
As previously described, these can be probed at low charge carrier concentrations 
($p = \unit[1 \times 10^{18}]{cm^{-3}}$, red solid lines in \f{thick}). 
As shown in \f{thick}(a) the behavior of the total electrical conductivity than differs evidently from the Fuchs-Sondheimer limit. 
At low film thickness we find enhanced values of $\nicefrac{\sigma_{\text{tot}}}{\sigma_{\infty}} \approx 12$. 
Obviously the contributions of the SS's dominate the transport and lift the total electrical conductivity above the bulk limit. 
As indicated in \f{thick}(a) the contribution of the SS to the total transport exceeds 90\% (cf. the thin red dashed lines). 
A comparable crossover between surface state-dominated and Fuchs-Sondheimer transport was 
reported for ultrathin copper films \cite{Fedorov:2007p14991}. Generally speaking, for film thicknesses above 18 QL thickness the $\bm{k}$-dependency 
and spatial distribution of the surface states is fully established. Thus, the surface contribution to the transport properties 
remain robust for thicker films and the TDF scales as 
$\Sigma_{SS}(d) \sim \nicefrac{1}{d}$. Quantum confinement effects are hardly visible for the 160 QL film, 
\textit{i.e.} $\sigma_{\text{bulk}}(d) \approx \sigma_{\infty}$ and $\sigma_{\text{tot}}(d) \approx \sigma_{\infty} + \nicefrac{\sigma_{\text{SS}}}{d}$.

The dependence of the normalized total thermopower on the film thickness is presented in \f{thick}(b) and (d). 
Keeping in mind the aforementioned 
relation $\nicefrac{S_{\text{tot}}}{S_{\infty}} \approx (\sigma_{\infty}+\sigma_{\text{SS}}\times\nicefrac{S_{\text{SS}}}{S_{\infty}})/\sigma_{\text{tot}}$, 
$\nicefrac{S_{\text{tot}}}{S_{\infty}} \le 1$ can be expected for a broad range of temperatures and doping in 
the thin TI films\footnote{We note that there is a difference of 
$S_{\text{bulk}}$ and $S_{\infty}$, which are the bulk contribution to a thin film and the contribution of a perfect infinite bulk, respectively. However, 
$S_{\text{bulk}}=S_{\infty}$ is used in this estimation and is valid for qualitative discussions.}. 
This is related to the mere fact, that in a TI the SS is metallic, hence showing small values of $|S_{\text{SS}}|$, while the bulk is 
insulating yielding large values of $|S_{\infty}|$. The position of the Dirac point near the band edge, \textit{i.e.} the case in \ce{Sb2Te3} and \ce{Bi2Te3}, 
leads moreover to opposite signs of $S_{\text{SS}}$ and $S_{\infty}$, further reducing $\nicefrac{S_{\text{tot}}}{S_{\infty}}$. 

Occasionally, the bulk and the SS's contribution have the same sign and $S_{\text{SS}} > S_{\infty}$, leading 
to an enlarged $\nicefrac{S_{\text{tot}}}{S_{\infty}}>1$. The latter scenario is seen in \f{temp}(f) for a heavily $p$-doped sample at temperatures lower than $T=\unit[180]{K}$. However, as these situations occur only if the bulk is metallic, 
\textit{i.e.} $\sigma_{\text{bulk}} \gg \sigma_{\text{SS}}$, the total thermopower will not exceed the bulk limit considerably, 
yielding $S_{\text{tot}} \approx S_{\infty}$. 

\subsection{Experimental results}
\begin{figure}[t!]
\centering
\includegraphics[width=0.35\textwidth]{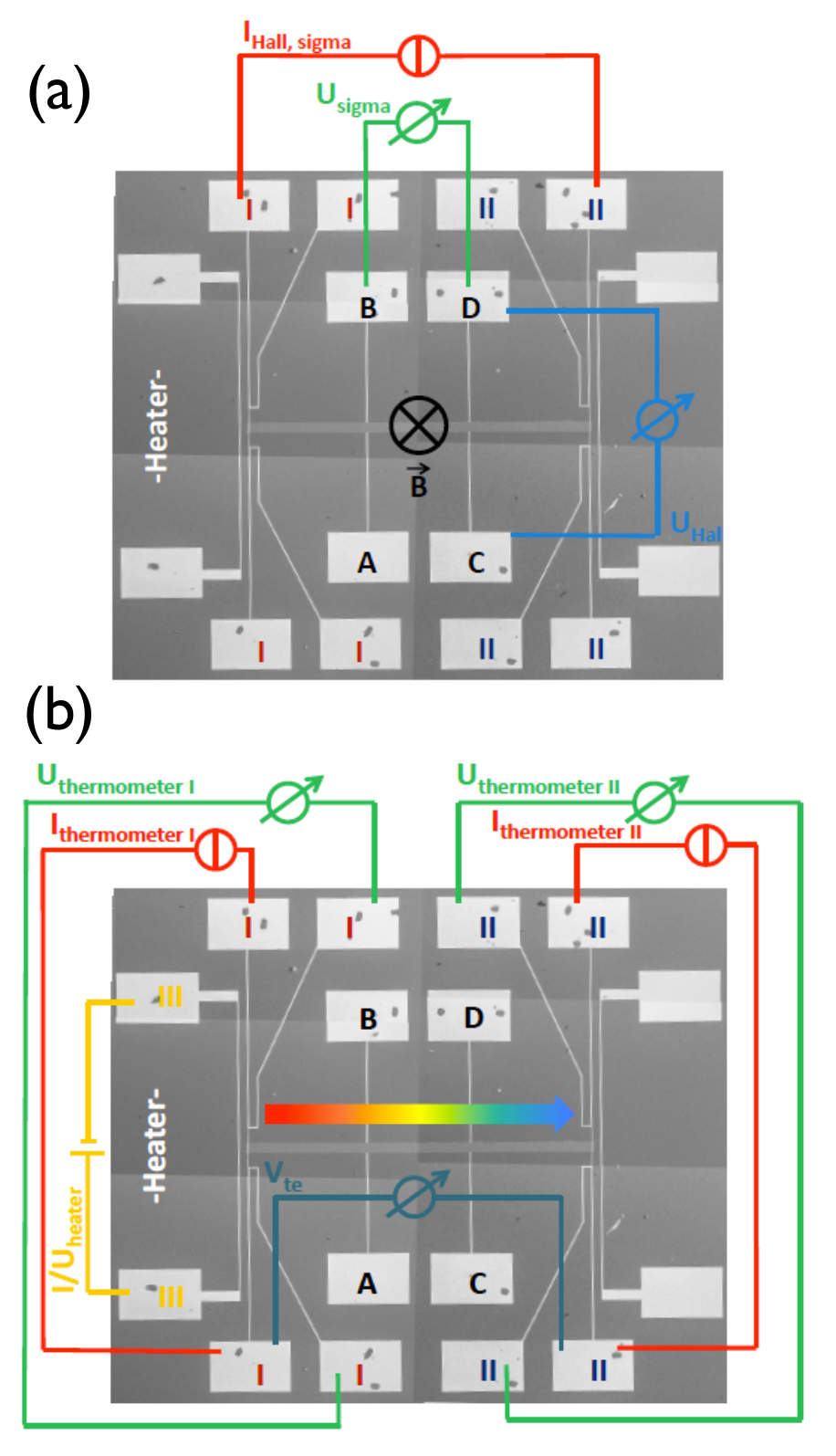}
  \caption{(color online) Device for electrical conductivity $\sigma$ and Hall measurements (a) as well as for determining the Seebeck coefficient $S = V_{th}/(T_{I} - T_{II})$ (b) of a Sb$_2$Te$_3$ thin film. The magnetic field B is applied perpendicular to the film plane. Bright lines are lithographically defined metal lines used as heater, as ohmic inner electrodes for electric measurements and as resistive thermometers, as indicated. (see methods for details)}
  \label{fig:device}
\end{figure}
\begin{figure}[t!]
\centering
\includegraphics[width=0.45\textwidth]{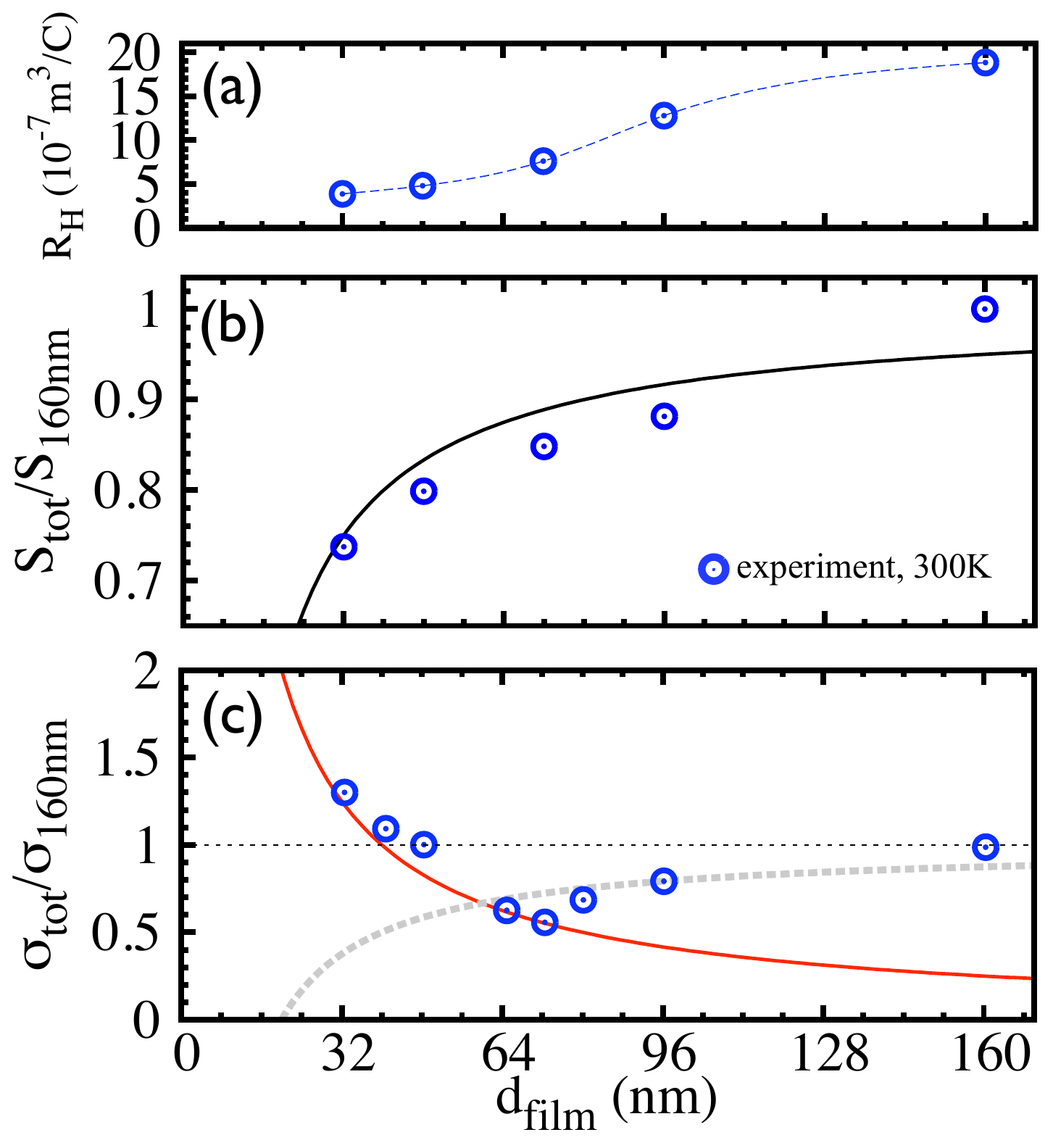}
  \caption{(color online) Measured thickness dependent thermoelectric transport properties at room temperature. Shown (as blue dots) is the normalized total thermopower (b) and the normalized total electrical conductivity (c) for varying film thickness. The charge carrier concentration varies monotonically with film thickness (a). Hence, the two transport regimes are superimposed in the experiment. The surface-state dominant regime (red solid line) and a Fuchs-Sondheimer transport regime (gray dashed line). The total thermopower (b) is always reduced due to the contribution of the surface-states.}
  \label{fig:thickexp}
\end{figure}

To support our theoretical findings at room temperature, experimental four-terminal measurements (see \f{device} and Methods) on the normalized total electrical conductivity and total thermopower at varying \ce{Sb2Te3} film thickness are presented in \fs{thickexp}. As can be seen from \f{thickexp}(b) the total thermopower tends 
asymptotically towards the bulk limit for increasing film thickness, \textit{i.e.} decreasing influence of the SSs, 
in accordance to the theoretical calculations in \f{thick}(b) (preferably red solid line). Obviously the measured Hall coefficient $R_H$ is not 
a constant with the film thickness (\f{thickexp}(a)). Within a two channel model for  $R_H$ \cite{Gonzalez:2014p16693,supp} 
and mobilities $\mu_{\text{SS}} \approx \mu_{\text{bulk}}$ \cite{Takagaki:2012p16691,supp} the charge carrier concentration $p$ 
of the thin films slightly varies monotonically with the number of deposited layers, too. 
That allows for direct observation of the transition between the surface-state-dominant and Fuchs-Sondheimer transport regime of the total electrical conductivity, shown in \f{thickexp}(c). 

At low film thickness $d<\unit[48]{nm}$ and still low enough charge carrier concentrations the SSs noticeably influence the transport, \textit{i.e.} the normalized total electrical conductivity is higher than expected from the bare bulk sample (indicated by red solid lines). 
With increasing film thickness the influence of the SSs is suppressed and bulk states dominate the transport. Nevertheless, the low charge carrier concentrations allow to see influences in the thermopower even at large film thicknesses, as pointed out in \f{thick}(b). 
Starting from a quantum mechanical approach a Fuchs-Sondheimer transport regime can be expected (indicated by gray dashed lines), with the normalized total electrical conductivity tending to the bulk value for large film thickness. Clearly the measured data supports qualitatively all aspects of the theoretical calculations, hence $\mu_{\text{SS}} \approx \mu_{\text{bulk}}$ is strongly expected.

To close our discussions we point to the low temperature case of $T=\unit[100]{K}$ shown in \f{thick}(b) and (d). 
While the main trends remain and the same conclusions as for the room-temperature case can be drawn, the influence of the SS on the total electrical 
conductivity being diminished (cf. \f{thick}(b)). Two facts lead to this result: (\textit{i}) the chemical potential is located deeper in the 
valence bands for a given charge carrier concentration (\textit{ii}) the broadening of the Fermi-Dirac distribution in 
Eq.\eqref{TDF} is reduced. Both facts lead to a reduced influence of the SSs located in the fundamental band gap. 
Deviations are only found for ultra-thin films of 3 QL thickness, for which the hybridization 
of the SSs at opposite sides of the film opens up gaps in the SS bands; at low temperature both the SS and the bulk interior behave then like a 
conventional semiconductor \cite{supp,Takahashi:2012,Osterhage:2014p16634}.

\section{Conclusion}
Combining quantum theoretical methods and experimental techniques on thin-film \ce{Sb2Te3} 
we reproduced and clarified the reduction the total thermopower found in various experiments on thin-film thermoelectric TIs\cite{Peranio:2006p15247,Zastrow:2014p16429,Boulouz:2001p16385}. 
By means of \textit{ab initio} electronic structure calculations based on density functional theory 
we discussed the thermoelectric properties of the $p$-type TI \ce{Sb2Te3} for various film thicknesses and temperatures. 
The topologically protected surface-state leads to metallic conduction of the thin films even in the
semiconducting regime. As shown by a separation of bulk and surface contributions, the latter leads 
to a strong reduction of the total measured thermopower of the thin films. 
This reduction is present in a wide range of film thicknesses and doping. By varying the charge carrier concentration a crossover between 
a surface-state-dominant and a Fuchs-Sondheimer transport regime is achieved. These two transport regimes, as well as a reduction 
of the total thermopower with decreasing film thickness are confirmed by thermoelectric transport measurements on atomic layer deposited 
\ce{Sb2Te3} thin films. To gain a thermoelectric benefit from thin film topological insulators, gapping the surface state at low temperatures 
and charge carrier concentrations seems to be the only favourable ansatz.



\section{Methods}
\textit{Computational details.}
The transport properties of the \SbTe films are calculated in the diffusive limit of transport by means of the semiclassical Boltzmann equation in relaxation time approximation (RTA) \cite{Mertig:1999p12776,Hinsche:2011p15707}. 
Within this approximation we assume that the attached heaters and metallic leads 
basically preserve the surface band structure.

The basis of the transport calculations, \textit{i.e.} the atomistic structure, was simulated 
by slabs of 15 to 800 atomic layers, \textit{i.e.} 3 -- 160 quintuple layers (QL) \ce{Sb2Te3}. The experimental in-plane lattice parameter 
${a^{\text{hex}}_{\text{SbTe}}}=4.264${\AA } and relaxed atomic positions \cite{Landolt} were used. 

The electronic structures of the \SbTe films were obtained by first principles calculations within density functional theory (DFT), 
as implemented in the \textsc{Quantum\-Espresso} code \cite{Giannozzi:2009p14969}. Fully relativistic, norm-conserving pseudo-potentials 
were used; exchange and correlation effects were accurately accounted for by the local density approximation (LDA) \cite{Perdew:1981p14970}. 

Subsequently the first-principles electronic structures were mapped onto tight-binding Hamiltonians 
\cite{Rauch:2013p16394,Rauch:2014p16553}. 
The resulting band structures were checked against our first-principles Korringa-Kohn-Rostoker 
\cite{Hinsche:2011p15707,Yavorsky:2011p15466} and \textsc{QuantumEspresso} results and yield fine agreement, 
in particular for the energy range near the fundamental band gap. 

The electronic structures serves as an input to obtain the thermoelectric transport properties, using the layer-resolved 
transport distribution function (TDF) $\Sigma_{i}(\mu)=\mathcal{L}_{i}^{(0)}(\mu, 0)$ 
~\cite{Mahan:1996p508}. The generalized conductance moments $\mathcal{L}_{i}^{(n)}(\mu, T)$ are defined as 
\begin{align}
\begin{split}
\mathcal{L}_{i}^{(n)}(\mu, T) &= \frac{\tau}{(2\pi)^2} \sum \limits_{\nu} \int d^2 \bm{k} \ |v^{\nu}_{\bm{k}}|^2 \cdot \label{TDF}
\\ 
&\mathcal{P}_{\bm{k}}^i \ (E^{\nu}_{\bm{k}} -\mu)^{n}\left( -\frac{\partial f(\mu,T)}{\partial E} \right)_{E=E^{\nu}_{\bm{k}}} . 
\end{split}
\end{align}
$v^{\nu}_{\bm{k}}$ denotes the group velocities in the directions of the hexagonal basal plane and $\mathcal{P}_{\bm{k}}^i$ is 
the layer-resolved probability amplitude of a Bloch state, which allows for spatial decomposition of the transport properties 
\cite{Zahn:1998p15803}. Details on this projection technique are published elsewhere \cite{Rittweger:2014p16422}.
%
The relaxation time for \SbTe was fitted to experimental data and chosen constant 
with absolute value $\tau=\unit[12]{fs}$ with respect to wave vector $\bm{k}$ and energy 
on the scale of $k_{B}T$ \cite{Hinsche:2011p15707}. 
The influence of electron-phonon coupling was theoretically and experimentally found to be very weak and is 
discussed in detail in Ref.~\citenum{Rittweger:2014p16422}. Assuming stoichiometric samples, 
exchange-defects of \ce{Sb} and \ce{Te1} are the most probable scattering centers to be expected \cite{Thonhauser:2003p15461}. 
Thus electron-impurity scattering will dominate. We checked the state-dependency of electron-impurity relaxation time $\tau_{\bm{k}}$ in 
Born approximation \cite{supp} and found a constant Brillouin zone averaged value of $\tau$ to be a reasonable approximation. 
The temperature- and doping-dependent in-plane 
electrical conductivity $\sigma$ and thermopower $S$ read 
\begin{equation}
\sigma=2e^2 \mathcal{L}^{(0)}(\mu, T) \quad \text{and} \quad S=\frac{1} {eT} \frac{\mathcal{L}^{(1)}(\mu,T)} {\mathcal{L}^{(0)}(\mu,T)}
\label{sigma}
\end{equation}
for given chemical potential $\mu$ at temperature $T$ and fixed extrinsic carrier concentration. 
%

The combination of first-principles and related tight-binding calculations allows for dense adaptive k-point meshes to ensure the convergence of Eq.~\eqref{TDF} \cite{Zahn:2011p15523,Yavorsky:2011p15466}. The calculation consists of more than 500 points in a piece of the 2D Fermi surface in the irreducible part of the Brillouin zone (BZ).

\textit{Experimental details.}
%
%
%
The \ce{Sb2Te3} thin films were grown on silicon wafers with a top layer of $\unit[300]{nm}$ \ce{SiO2} \textit{via} Atomic Layer Deposition (ALD) 
at substrate temperatures of $\unit[353]{K}$. \ce{(Et3Si)2Te} and \ce{SbCl3} were used as precursors at 
source temperatures of $\unit[350]{K}$ and $\unit[328]{K}$, respectively \cite{Pore:2009p16689}.
For the transport measurements the \ce{Sb2Te3} thin films were deposited on a lithographically pre-patterned Hall bar, which was defined \textit{via} laser beam lithography and subsequent developing of the exposed photoresist. Contacts to the hall bars were defined in a second lithography step, prior to sputter deposition of Ti/Pt metal contacts. The thickness t and lateral dimensions (length l, width w) of the films were obtained by Atomic Force Microscopy and SEM images, respectively (cf. supplemental material). 
The Hall resistance $R_H = U_{C,D}/I_{I,II}$ of the films was determined with standard lock-in technique. A constant ac current with an amplitude of 
$I_{I,II} =  \unit[10]{\mu A}$ and frequency of $\unit[6]{Hz}$ was applied along the film stripe between contacts $I$ and $II$ and the voltage drop 
$U_{C,D}$ across the film width has been measured  between contacts C and D, while sweeping a magnetic field 
from $\unit[-3]{}$ to $\unit[3]{T}$ in $\unit[0.1]{T}$ steps (cf. \f{device}(a)). The magnetic field has been applied perpendicular to the film plane. 
The electrical conductivity has been calculated using $\sigma =\left[l/(wt) \right]/R_H $ where $R_H = U_{C,D}/I_{I,II}$ 
is the resistance of the film, measured in a four-point configuration between contacts B and D. 
To determine the Seebeck coefficient $S = V_{th}/\Delta T$ of the films the on-chip line heater generated a temperature difference $\Delta T$ across its length. $\Delta T$ was measured by resistance thermometry using two metal four-probe thermometer lines ($I$ and $II$) located at the film  ends (cf. \f{device}(b)), which were driven by standard lock-in technique in a four-terminal configuration \cite{Bae:2014p16688,Zastrow:2014p16429}. The metal lines for thermometry also served as electrodes for measuring $V_{th}$. Expected error bars for film thickness, thermopower and electrical conductivity are $\sim$10\%, $\sim$16\% and $\sim$7\%, respectively.

\begin{acknowledgement}
The authors acknowledge financial support by the Priority Programs SPP 1386 and SPP 1666 of DFG\@.
\end{acknowledgement}

\begin{suppinfo}
Additional results on the doping- and temperature-dependent electrical conductivity and 
thermopower for other film thicknesses, informations on the electronic structure at 
varying film thickness, as well as structural analysis of the ALD grown thin films are available on-line.
\end{suppinfo}

\bibliography{nl_0115_ST.bbl}

 \onecolumn
 \setcounter{figure}{0}
 
\section*{Impact of the Topological Surface State on the Thermoelectric Transport in Sb$_2$Te$_3$ Thin Films: Supplemental Material} 
 
 \begin{figure*}[h!]
\centering
\includegraphics[width=0.89\textwidth]{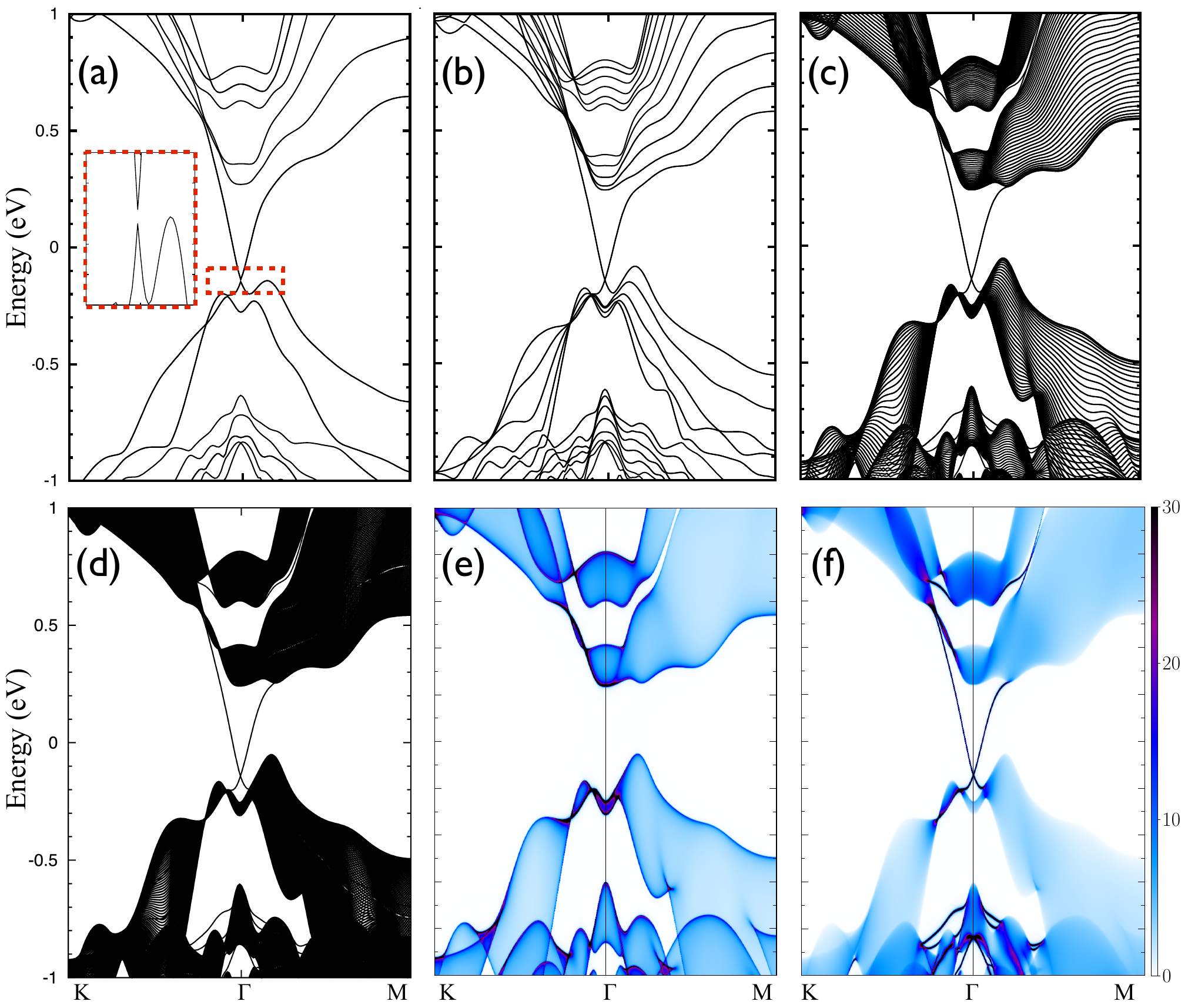}
  \caption{(color online) \textit{Ab initio} based tight-binding band structures of (a) 3 QL, (b) 6QL, (c) 36QL and (d) 160QL thick $\text{Sb}_2\text{Te}_3$ films along the high symmetry lines of the two dimensional Brillouin zone. The inset in (a) highlights the hybridization-gapped Dirac state. Additionaly, the spectral function for (e) an infinite sample and (f) a semi-infinite sample are shown.}
  \label{fig:bands}
\end{figure*}

Being the basis of our calculations, exemplary bandstructures of 3QL, 6QL,  36QL and 160QL thick \ce{Sb2Te3} 
films are shown in \fs{bands}(a)-(d), respectively. 1QL (quintuple layer), comprises five atomic layers and a 
thickness of $\unit[1.02]{nm}$. The prototypical linear dispersion of the SSs 
in the fundamental band gap and warping at elevated energies towards the conduction band edge is already well reproduced for very 
thin films of 3QL (\f{bands}(a)), while being slightly gapped due to the hybridization of the SSs on the upper and 
lower side of the film (cf. inset of \f{bands}(a)). For increasing film thickness the fundamental band gap $E_g$ decreases ($E_{g,3QL}\approx\unit[413]{meV}$;$E_{g,6QL}\approx\unit[329]{meV}$;$E_{g,36QL}\approx\unit[295]{meV}$) and quantum confinement effects bury the Dirac point 
deeper below the valence band edge (cf. \f{gap}). With \ce{Sb2Te3} being an inherently p-doped material, ultra-thin films could therefore provide an opportunity to probe the exposed Dirac point, like in \ce{Bi2Se3} but in contrast to \ce{Bi2Te3}. 

\begin{figure*}[t]
\centering
\includegraphics[width=0.7\textwidth]{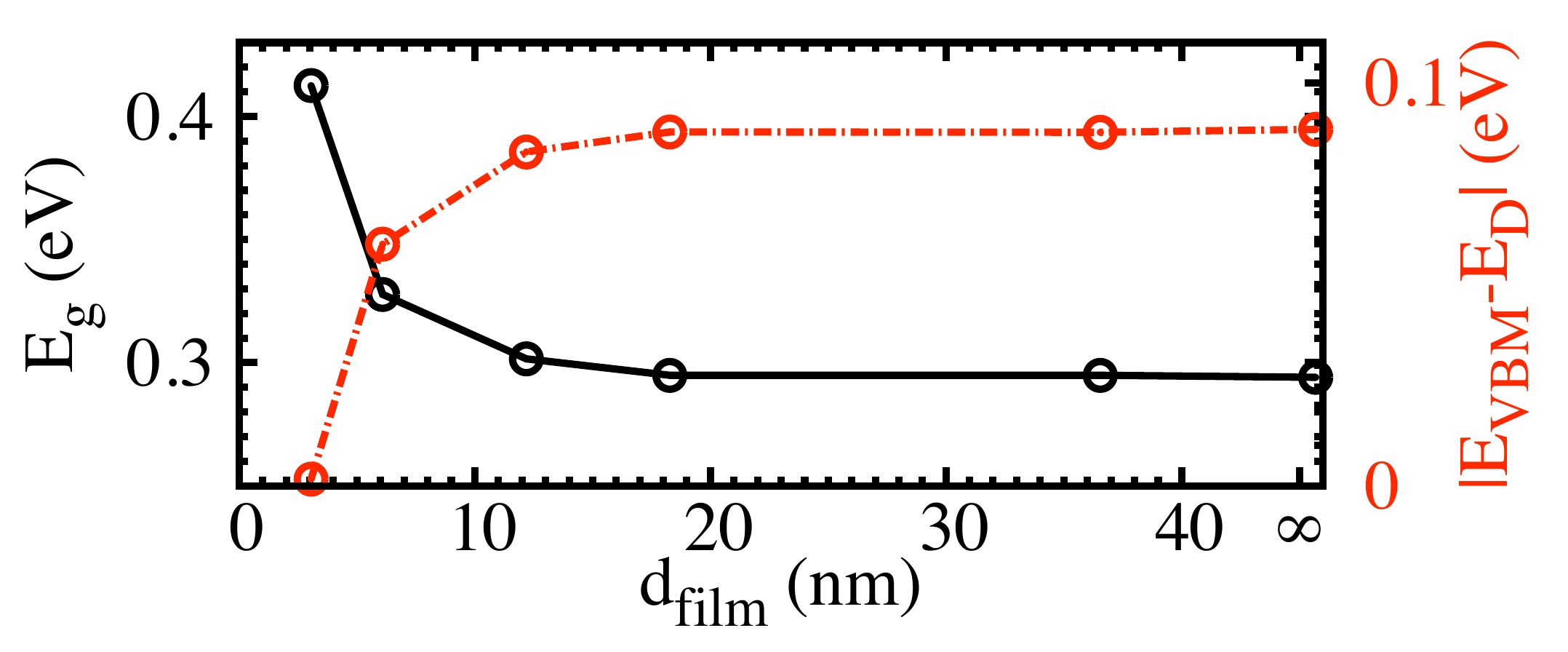}
  \caption{(color online) Calculated evolution of the fundamental band gap of the \ce{Sb2Te3} thin films (left axis, black solid line) and the shift of the 
  Dirac point with respect to the valence band edge (right axis, red dash-dotted line). Infinity indicates calculations for semi-infinite films.}
    \label{fig:gap}
\end{figure*}

We show as in the main manuscript, the calculated electrical conductivity and thermopower 
in dependence on temperature for three distinct hole concentrations at the investigated film thicknesses of 3QL, 6QL, 12QL, 36QL and 160QL. 
The discussion can be done analogously to the case of 18QL. Deviations show up for the case of 3QL, for which the 
influence of the hybridization-gapped surface state at low temperature comes into play and the bulk and surface 
are semiconducting for small carrier concentrations. However, even in this regime of temperature and thickness the thermoelectric efficiency will only 
reach values exceeding the bulk limit for $T<\unit[50]{K}$. 

\begin{figure*}[t]
\centering
\includegraphics[width=0.7\textwidth]{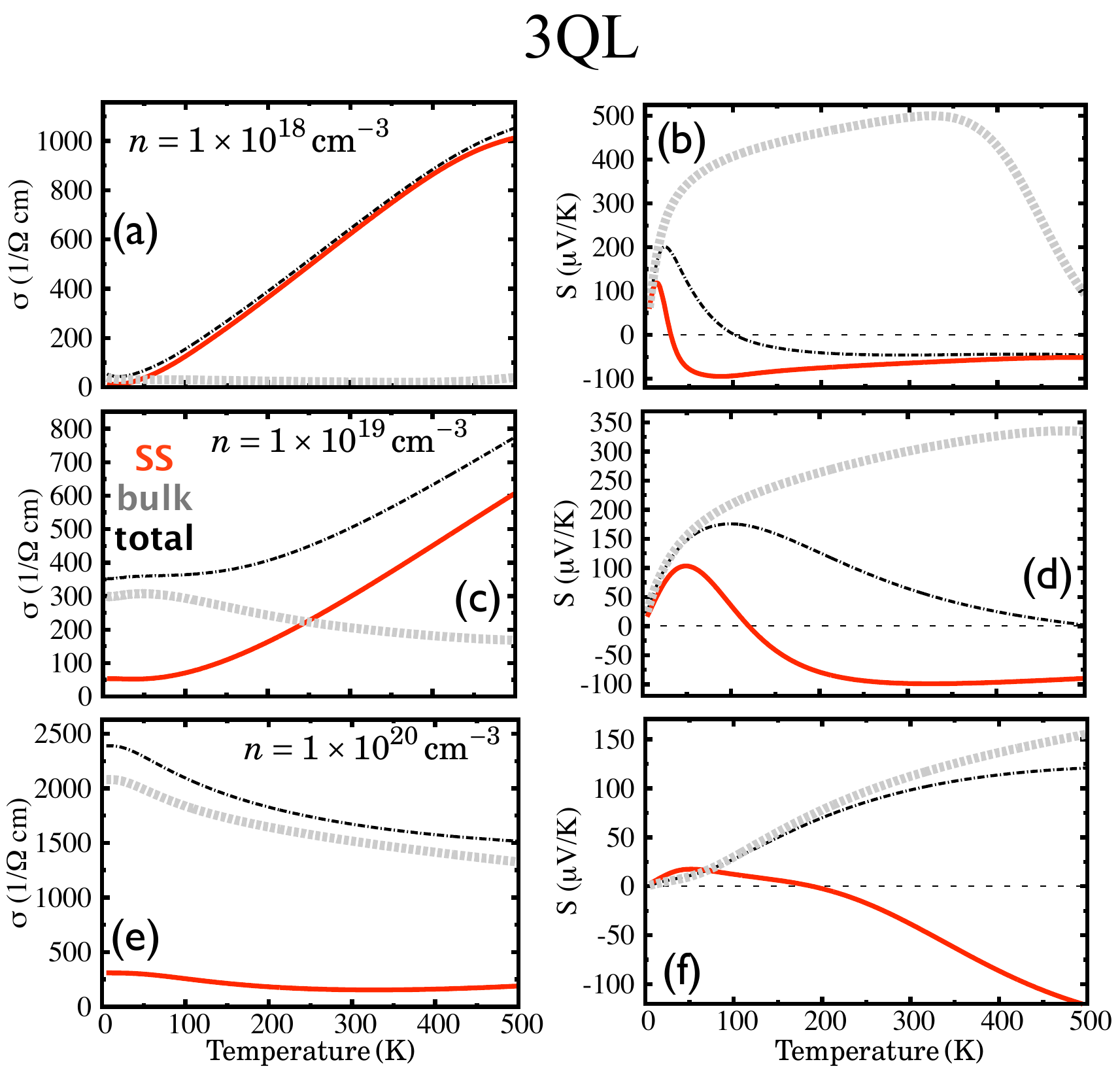}
  \caption{(color online) Calculated electrical conductivity (a),(c),(e) and thermopower (b),(d),(f) in dependence on temperature for three distinct 
hole concentrations at a film thickness of 3QL. 
(a) and (b) hole concentration of $n = \unit[1 \times 10^{18}]{cm^{-3}}$, (c) and (d) 
hole concentration of $n = \unit[1 \times 10^{19}]{cm^{-3}}$, (e) and (f) hole concentration 
of $n = \unit[1 \times 10^{20}]{cm^{-3}}$. Pure bulk contributions are stated by gray dashed lines, the contribution of the surface 
states is given by red solid lines, while black dash-dotted lines show the total contribution to the sample.}
\end{figure*}

\begin{figure*}[t]
\centering
\includegraphics[width=0.7\textwidth]{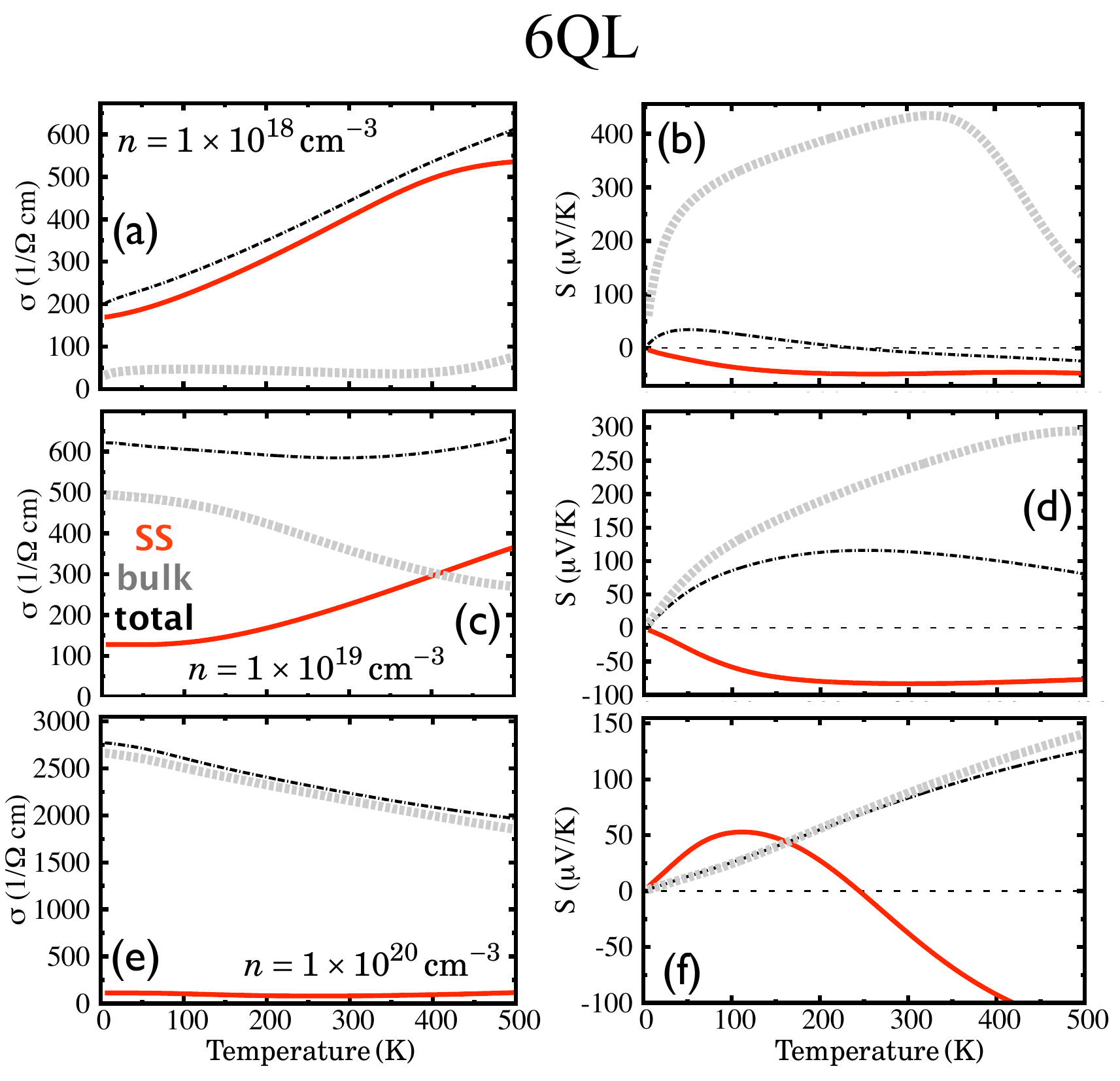}
  \caption{(color online) As Fig. 3 but for 6QL.}
\end{figure*}

\begin{figure*}[t]
\centering
\includegraphics[width=0.7\textwidth]{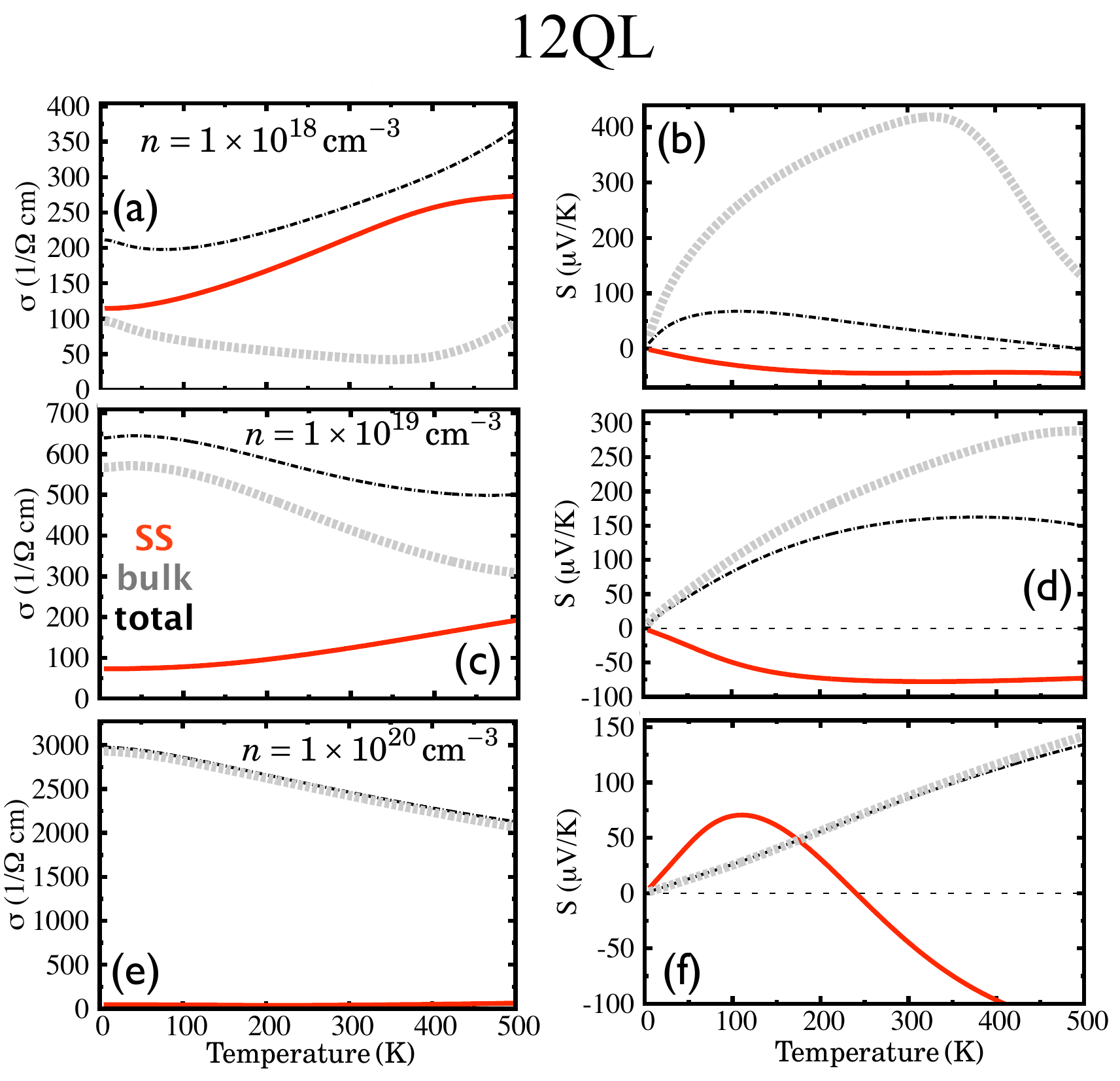}
  \caption{(color online) As Fig. 3 but for 12QL.}
\end{figure*}

\begin{figure*}[t]
\centering
\includegraphics[width=0.7\textwidth]{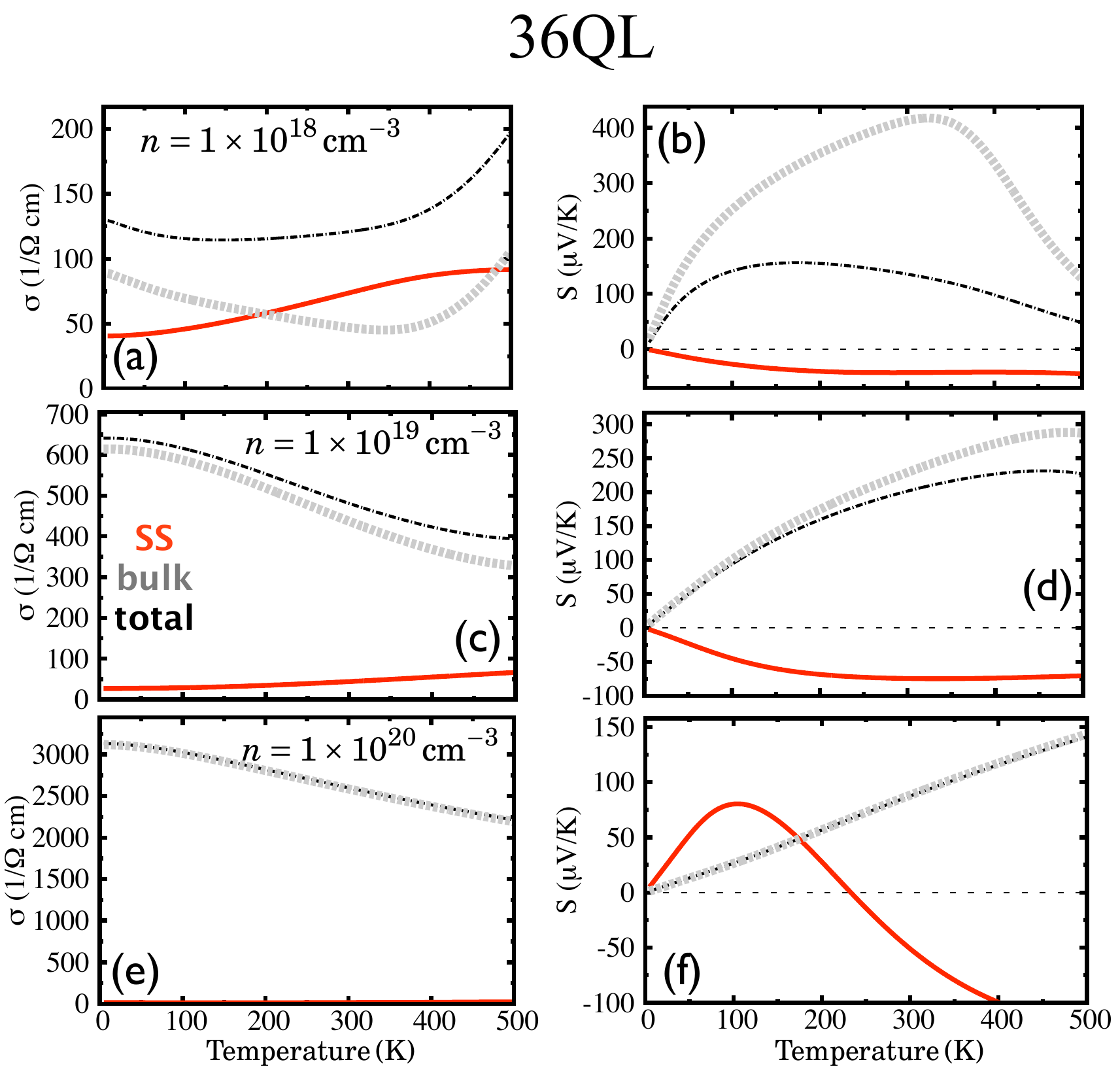}
  \caption{(color online) As Fig. 3 but for 36QL.}
\end{figure*}

\begin{figure*}[t]
\centering
\includegraphics[width=0.7\textwidth]{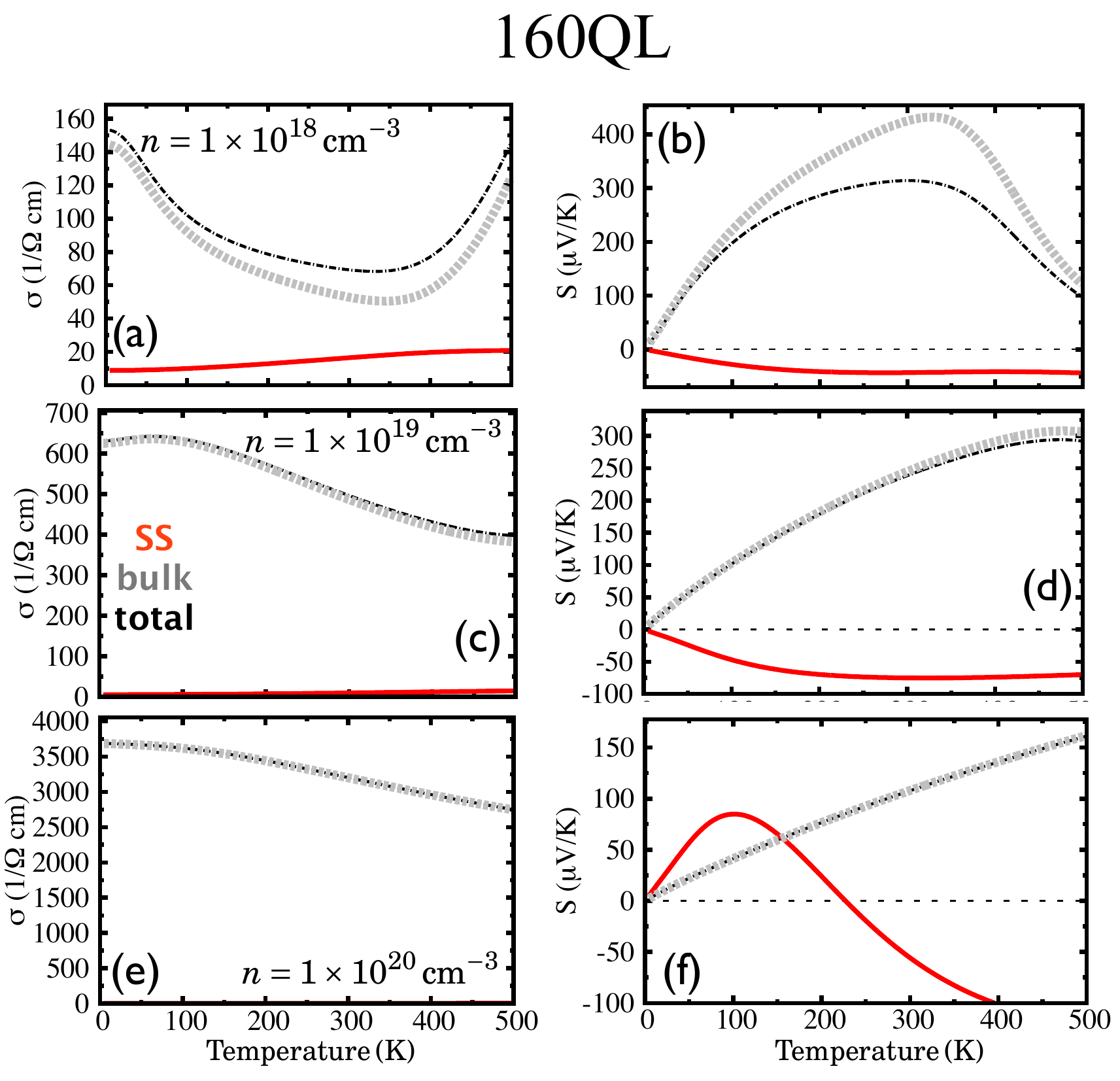}
  \caption{(color online) As Fig. 3 but for 160QL.}
\end{figure*}

In figure 8 the dependence of the charge carrier concentration of the surface state, the bulk interior and the total sample are shown in dependence of the chemical potential at room temperature. We emphasize that for p-doped \ce{Sb2Te3} a correspondence for the \textsc{Hall} conductivity within an one-channel
\begin{equation}
R_{H} = 1/ (e \times n_{\text{tot}})
\end{equation}
and a two-channel model
\begin{equation}
R_{H} = (n_{\text{SS}} \times \mu_{\text{SS}}^2 + n_{\text{bulk}} \times \mu_{\text{bulk}}^2) / (e \times (n_{\text{SS}} \times \mu_{\text{SS}} + n_{\text{bulk}} \times \mu_{\text{bulk}})^2)
\end{equation}
can only be found if $\mu_{\text{SS}} \approx \mu_{\text{bulk}}$. Our theoretical calculations allow for the determination of $n_{\text{SS}}$, $n_{\text{bulk}}$ and $n_{\text{tot}}$. Calculating $R_{H}$ within the one- and two-channel model and comparing to the experimental 
results strongly suggests $\mu_{\text{SS}} \approx \mu_{\text{bulk}}$. An enhanced mobility $\mu_{\text{SS}} \gg \mu_{\text{bulk}}$ of the 
surface states is only to be expected for a non-hybridized, then back-scattering free \textsc{Dirac} cone, which could be the case for weakly \textit{n}-doped \ce{Sb2Te3}.

\begin{figure*}[h]
\centering
\includegraphics[width=0.7\textwidth]{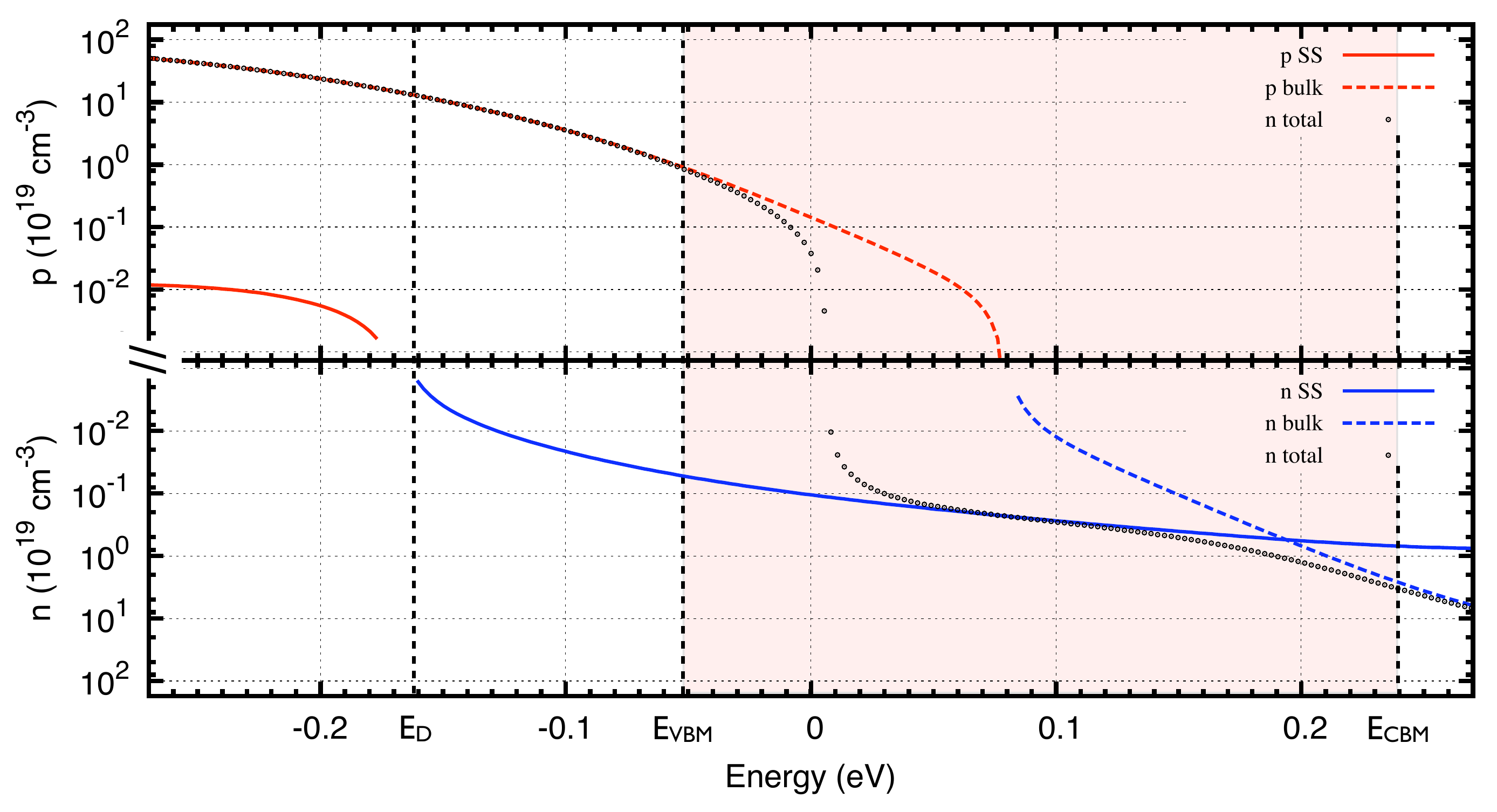}
  \caption{(color online) Calculated electron (n, bottom) and hole (p, top) concentration as position of the chemical potential at room temperature for a 36QL thin film. Shown are the partial contribution from the bulk bands (dashed lines), the surface state (solid lines) and the total carrier concentration (dots) assuming comparable mobilities $\mu_{\text{SS}} \approx \mu_{\text{bulk}}$. The fundamental band gap is indicated by the shaded area. For  $p$-type thin films the contribution of the SS to the charge carrier concentration plays a minor role.}
\end{figure*}

To support the used constant relaxation time $\tau$ we carried out calculations of the state-dependent electron-impurity relaxation time $\tau_{\bm{k}}$ in Born approximation as previously introduced in 'Zahn \textit{et al.}, Phys. Rev. Lett. 1998, 80, 194309'. Assuming stoichiometric samples, exchange-defects of \ce{Sb} and \ce{Te1} are the most probable scattering centers to be expected. The relaxation time $\tau_{\bm{k}}$ for $p$-doped bulk \ce{Sb2Te3} with a hole concentration of about $n = \unit[1 \times 10^{19}]{cm^{-3}}$ is shown in Fig. 9. While, some state-dependency of the relaxation time can be found, using an Brillouin zone averaged value of $\langle \tau_{\bm{k}} \rangle = \unit[12]{fs}$ seems to be a reliable approximation. The energy-dependence of $\tau_{\bm{k}}$ is rather weak in the range of the investigated charge carrier concentrations. 
\begin{figure*}[h]
\centering
\includegraphics[width=0.7\textwidth]{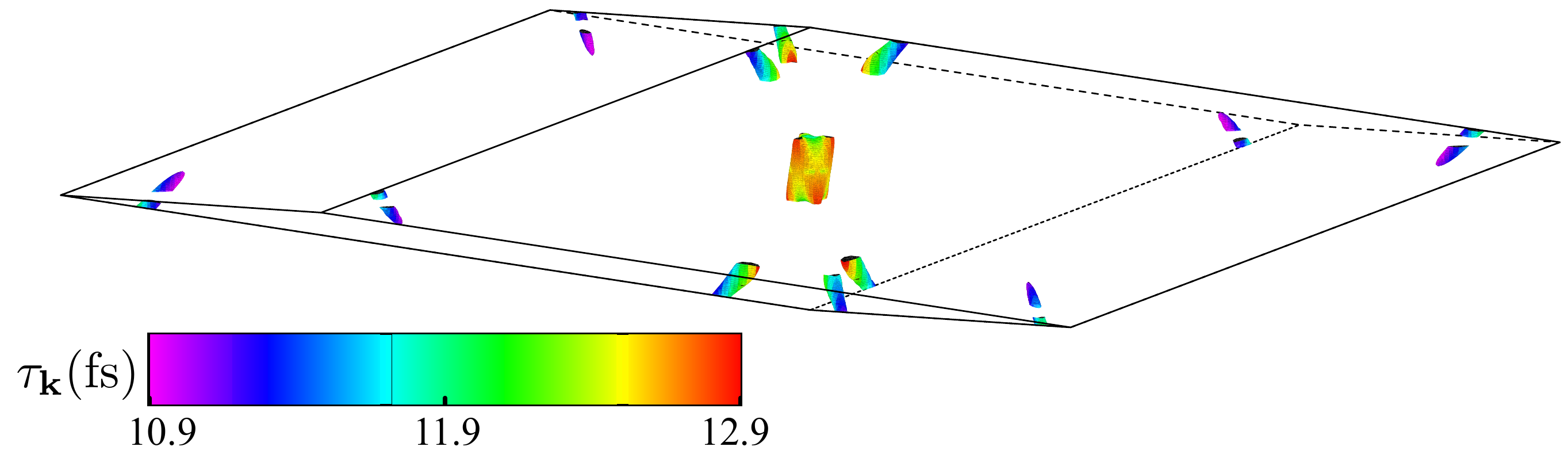}
  \caption{(color online) Calculated state-dependent electron-impurity relaxation time $\tau_{\bm{k}}$ in Born approximation for $p$-doped bulk \ce{Sb2Te3} with a hole concentration of about $n = \unit[1 \times 10^{19}]{cm^{-3}}$. Exchange-defects of \ce{Sb} and \ce{Te1} are assumed to be the main scattering centers.}
\end{figure*}

In figures 10 and 11 experimental x-ray-diffraction (XRD) and scanning-electron-microscopic (SEM) data of the ALD grown thin films are shown, 
to depict the experimental quality and reliability of the underlying material system. 
More in-depth analysis and discussion of the experimental methods can be found in 'Zastrow \textit{et al.}, 
Semiconductor Science and Technology 2013, 28, 035010' and 'Bae \textit{et al.}, Semiconductor Science and Technology 2014, 29, 064003'.
\begin{figure*}[h]
\centering
\includegraphics[width=0.7\textwidth]{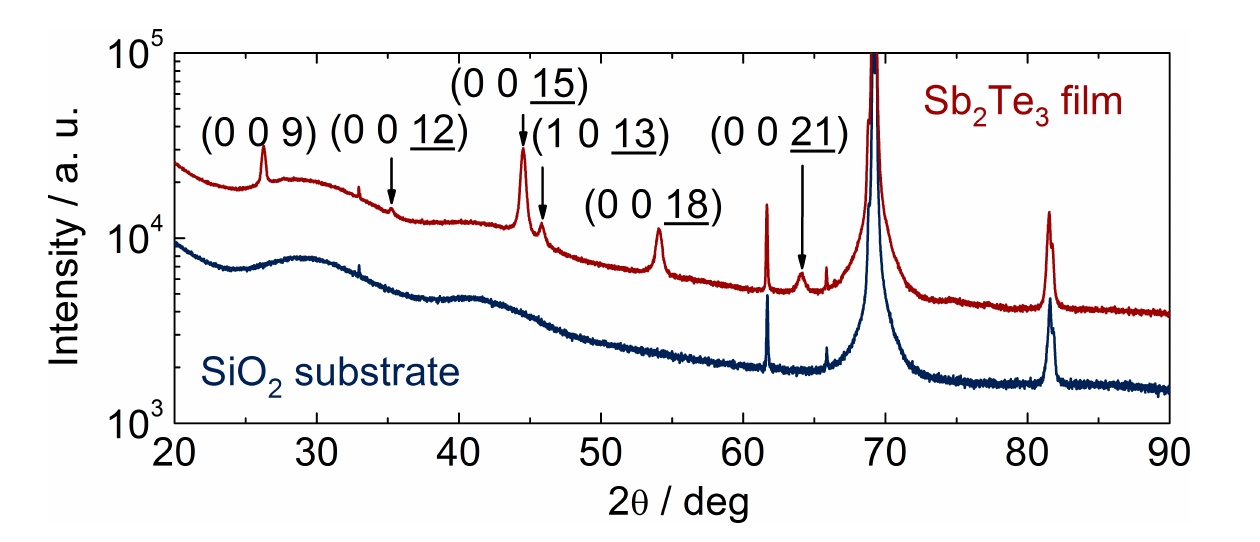}
  \caption{(color online) Measured XRD-pattern of a atomic layer-deposited (ALD) 50nm \ce{Sb2Te3} thin film and of the \ce{Si}/\ce{SiO2} substrate.}
\end{figure*}

\begin{figure*}[h!]
\centering
\includegraphics[width=0.55\textwidth]{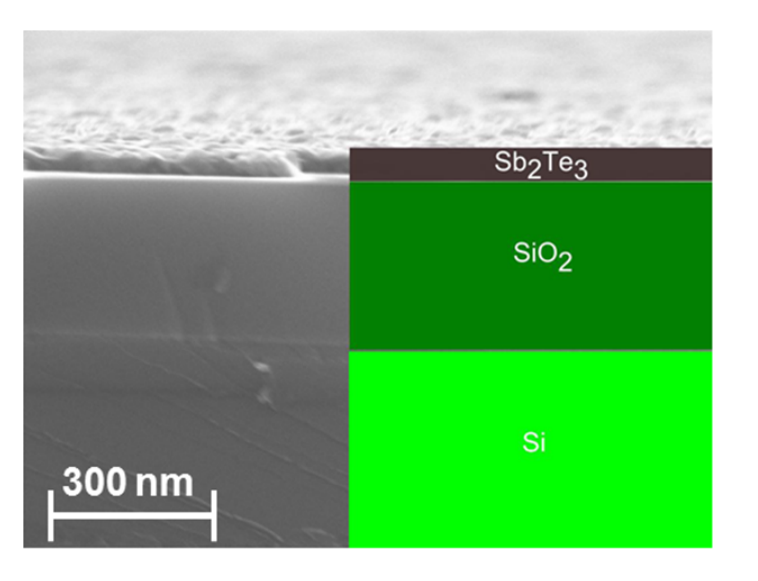}
  \caption{(color online) SEM cut view of a ALD \ce{Sb2Te3} thin film grown on top of a \ce{Si}/\ce{SiO2} wafer.}
\end{figure*}

\end{document}